\documentstyle[aps]{revtex}
\begin{document}
\def\Universita{Universit\'a}
\def\Paris{Par\'{\i}s}
\def\Perez{P\'erez}
\def\Gunther{G\"unther}
\def\Schutzhold{Sch\"utzhold}
\def\Lofstedt{L\"{o}fstedt}
\def\Garcia{Garc\'{\i}a}
\def\ni{n_{\mathrm in}}
\def\no{n_{\mathrm out}}
\def\half{{\textstyle{1\over2}}}
\def\in{{\mathrm in}}
\def\out{{\mathrm out}}
\title{\bf Sonoluminescence: \\
\bf Two-photon correlations as a test of thermality} 
\author{F. Belgiorno$^{a}$}
\address{Universit\`a degli Studi di Milano, Dipartimento di Fisica, 
Via Celoria 16, 20133 Milano, Italy,\\ 
and \\
I.N.F.N., Sezione di Milano, Italy}
\author{S. Liberati$^{b}$} 
\address{International School for Advanced Studies, 
Via Beirut 2-4, 34014 Trieste, Italy}
\author{Matt Visser$^{c}$} 
\address{Physics Department, Washington University, 
Saint Louis MO 63130-4899, USA}
\author{D.W. Sciama}
\address{International School for Advanced Studies, 
Via Beirut 2-4, 34014 Trieste, Italy, \\
and\\
International Center for Theoretical Physics,  
Strada Costiera 11, 34014 Trieste, Italy, \\
and\\
Physics Department, Oxford University, Oxford, England}

\date{{\sf quant-ph/9904018}; Revised 27 May 2000; \LaTeX-ed \today}
\maketitle
\begin{abstract}

In this Letter we propose a fundamental test for probing the
thermal nature of the spectrum emitted by sonoluminescence.  We show
that two-photon correlations can in principle discriminate between
real thermal light and the pseudo-thermal squeezed-state photons
typical of models based on the dynamic Casimir effect. Two-photon
correlations provide a powerful experimental test for various classes
of sonoluminescence models.

\end{abstract}

To appear in {\em Physics Letters {\bf A}.}

\section{Introduction}

In this Letter we propose a fundamental test for experimentally
discriminating between various classes of theoretical models for
sonoluminescence.  It is well known that the optical photons measured
in sonoluminescence are characterized by a broadband spectrum, often
described as approximately thermal with a ``temperature'' of several
tens of thousands of Kelvin~\cite{Physics-Reports}.  Whether or not
this ``temperature'' represents an actual thermal ensemble is less
than clear. For instance, according to the ``shock wave approach'' of
Barber, Putterman {\em et al.}, or the ``adiabatic heating
hypothesis'', thermality of the spectrum is due to a high physical
temperature caused by compression of the gases contained in the
bubble. On the other hand, in models based on variants of Schwinger's
``dynamical Casimir approach''~\cite{Schwinger,Laeff,Los-Alamos}, it
is possible to avoid reaching high physical temperatures and yet to
obtain a thermal spectrum (or at least pseudo-thermal characteristics
for the emitted photons) because of the peculiar statistical
properties of the two-photon squeezed-states produced by this class of
mechanism.

We stress that thermal characteristics in single photon measurements
can be associated with {\em at least} two hypotheses: (a) real
physical thermalization of the photon environment; (b) pseudo-thermal
single photon statistics due to tracing over the unobserved member of
a photon pair that is actually produced in a two-mode squeezed state.
We shall call case (a) {\em real thermality}; while case (b) will be
denoted {\em effective thermality}.  Of course, case (b) has no
relation with any concept of thermodynamic temperature, though to any
such squeezed state one may assign a (possibly mode-dependent) {\em
effective temperature}.

Our aim is to find a class of measurements able to discriminate
between cases (a) and (b), and to understand the origin of the roughly
thermal spectrum for sonoluminescence in the visible frequency
range. In principle, the thermal character of the experimental
spectrum could disappear at higher frequencies, but for such
frequencies the water medium is opaque, and it is not clear how we
could detect them.  (Except through heating effects.) Our key remark
is that it is not necessary to try to measure higher than visible
frequencies in order to get a definitive answer regarding thermality.
It is sufficient, at least in principle, to measure photon pair
correlations in the visible portion of the sonoluminescence spectrum.
Thus regardless of the underlying mechanism, two-photon correlation
measurements are a very useful tool for discriminating between broad
classes of theory and thereby investigating the nature of
sonoluminescence.  We note that two-photon correlations have already
been proposed, for the first time in~\cite{Trentalange} and
subsequently in~\cite{HKP,SH}, as an efficient tool for measuring the
shape and the size of the emission region.  It was proposed
in~\cite{Trentalange,HKP} that precise Hanbury--Brown--Twiss
interferometry measurements could in principle distinguish between
chaotic (thermal) light emerging from a hot bubble and the possible
production of coherent light via the dynamical Casimir effect.
Unfortunately in the dynamical Casimir effect photons are always
pair-produced from the vacuum in two--mode squeezed states, not in
coherent states. Pair-production via the dynamical Casimir effect
appears to imply that all the photon pairs form two--mode squeezed
states, which are very different from the coherent states analyzed
in~\cite{Trentalange,HKP,SH}.

\section{Real thermal light versus two-mode squeezed states}

The quantum optics mechanism that simulates a thermal spectrum [case
(b)] is based on two-mode squeezed-states defined by
\begin{equation} 
|\zeta_{ab}\rangle =\hbox{e}^{- \zeta (a^{\dagger}
b^{\dagger} - b a)} |0_{a},0_{b} \rangle, 
\end{equation} 
where $\zeta$ is (for our purposes) a real parameter though more
generally it can be chosen to be complex \cite{bk}. In quantum optics
a two-mode squeezed-state is typically associated with a so called
non-degenerate parametric amplifier (one of the two photons is called
``signal'' and the other ``idler'' \cite{bk,bk2,yupo}). Consider the
operator algebra
\begin{equation}
[a,a^{\dagger}]=1=[b,b^{\dagger}],
\qquad
[a,b]=0=[a^{\dagger},b^{\dagger}],
\end{equation}
and the corresponding vacua 
\begin{equation}
|0_a \rangle :\ a |0_a \rangle =0, 
\qquad
|0_b \rangle :\ b |0_b \rangle =0. 
\end{equation}
The two-mode vacuum is the state $| \zeta \rangle \equiv
|0(\zeta) \rangle $ annihilated by the operators 
\begin{equation}
A (\zeta)=\cosh (\zeta) \; a -\sinh (\zeta) \; b^{\dagger},
\end{equation}
\begin{equation}
B (\zeta)=\cosh (\zeta) \; b-\sinh (\zeta) \; a^{\dagger}.
\end{equation}
A characteristic of two-mode squeezed states is that if we measure
only one photon and ``trace away'' the second, a thermal density
matrix is obtained \cite{bk,bk2,yupo}.  Indeed, if $O_a$ represents an
observable relative to one mode (say mode ``a'') its expectation value
on the squeezed vacuum is given by
\begin{equation}
\langle\zeta_{ab}|O_a|\zeta_{ab}\rangle
=\frac{1}{\cosh^2 (\zeta)} \;
\sum_{n=0}^{\infty} 
[\tanh(\zeta)]^{2 n} \; \langle n_{a}|O_a|n_{a} \rangle.    
\label{E:termo}
\end{equation}
In particular, if we consider $O_a=N_a$, the number operator in mode
$a$, the above reduces to
\begin{equation}
\langle\zeta_{ab}|N_{a}|\zeta_{ab}\rangle = \sinh^2 (\zeta).
\end{equation}
These formulae have a strong formal analogy with thermofield dynamics
(TFD) \cite{takume,ume}, where a doubling of the physical Hilbert
space of states is invoked in order to be able to rewrite the usual
Gibbs (mixed state) thermal average of an observable as an expectation
value with respect to a temperature-dependent ``vacuum'' state (the
thermofield vacuum, a pure state in the doubled Hilbert
space). In the TFD approach, a trace over the unphysical (fictitious)
states of the fictitious Hilbert space gives rise to thermal averages
for physical observables, completely analogous to the averages
in equation (\ref{E:termo}) {\em except} that we must make the
following identification
\begin{equation}
\tanh(\zeta)= \exp\left(-\frac{1}{2} \frac{\hbar \omega}{k_{B} T}\right),
\end{equation}
where $\omega$ is the mode frequency and $T$ is the temperature.  We
note that the above identification implies that the squeezing
parameter $\zeta$ in TFD is $\omega$-dependent in a very special way.

The formal analogy with TFD allows us to conclude that, provided 
we measure only one photon mode, the two-mode squeezed-state acts as a
thermofield vacuum and the single-mode expectation values acquire a
pseudo-thermal character corresponding to a ``temperature''
$T_{\mathrm squeezing}$ related with the squeezing parameter $\zeta$
by
\cite{yupo}
\begin{equation}
k_{B}\; T_{\mathrm squeezing} =
\frac{\hbar \; \omega_i}{2 \log(\coth(\zeta))},
\end{equation}
where the index $i=a,b$ indicates the signal mode or the idler mode
respectively; note that ``signal'' and ``idler'' modes can have
different effective temperatures (in general $\omega_{signal} \neq
\omega_{idler}$)~\cite{yupo}.

\section{A toy model and sonoluminescence}

To treat sonoluminescence, we introduce a quantum field theory
characterized by an infinite set of bosonic oscillators (as in bosonic
TFD; not just two oscillators as in the case of ``signal-idler''
systems studied in quantum optics).  The simple two-mode squeezed
vacuum is replaced by
\begin{eqnarray}
\label{E:general}
&&|\Omega[\zeta(k,k')]\rangle \equiv 
\exp\left[
-\int d^3 k \; d^3k' \;  \zeta(k,k') \;
(a_{k} b_{k'}-a^{\dagger}_{k} b^{\dagger}_{k'})\right] \Big|0\Big\rangle,
\end{eqnarray}
where the function $\zeta(k,k')$ is peaked near $k+k'=0$, and becomes
proportional to a delta function in the case of infinite volume
[$\zeta(k,k') \to \zeta(k) \delta(k+k')$] when the photons are emitted
strictly back-to-back~\cite{SL-prl,QED0,QED1,QED2}.  To be concrete,
let us refer to the homogeneous dielectric model presented
in~\cite{QED1}. In this limit there is no ``mixing'' and everything
reduces to a sum of two-mode squeezed-states, where each pair of
back-to-back modes is decoupled from the other. The frequency $\omega$
is the same for each photon in the couple, in such a way that we are
sure to get the same ``temperature'' for both. The two-mode squeezed
vacuum then simplifies to
\begin{equation}
|\Omega (\zeta_k)\rangle \equiv 
\exp\left[-\int d^3 k\; \zeta_k \;
(a_k a_{-k}-a^{\dagger}_k a^{\dagger}_{-k})\right] \Big|0\Big\rangle.
\end{equation}
The key to the present proposal is that, if photons are pair produced
via the dynamical Casimir effect, then they are actually produced in
some combination of these two-mode
squeezed-states~\cite{SL-prl,QED0,QED1,QED2}.  In this case
$T_{\mathrm squeezing}$ is a function of both frequency and squeezing
parameter, and in general only a special ``fine tuning'' would allow
us to get the {\em same} effective temperature for all couples. If we
consider the expectation value on the state $|\Omega (\zeta_k) \rangle
$ of $N_k\equiv a^{\dagger}_k a_k$ we get
\begin{equation}
\langle \Omega(\zeta_k)|N_k| \Omega(\zeta_k)\rangle =\sinh^2 (\zeta_k),
\end{equation}
so we again find a ``thermal'' distribution for each value of $k$
with temperature
\begin{equation}
k_{B}T_k\equiv \frac{\hbar\omega_{k}}{2\;\log(\coth(\zeta_k))}.
\end{equation}
The point is that for $k \neq {\bar{k}}$ we generally get $T_k \neq
T_{\bar{k}}$ {\em unless a fine tuning condition holds}.  This
condition is implicitly enforced in the definition of the
thermofield vacuum and it is possible only if we have
\begin{equation}
\coth (\zeta_k)=\hbox{e}^{ \kappa \omega_k},
\label{finet}
\end{equation}
with $\kappa$ some constant, so that the frequency dependence in
$T_k$ is canceled and the same $T_{\mathrm squeezing}$ is obtained for
all couples.

For models of sonoluminescence based on the dynamical Casimir effect
({\em i.e.} squeezing the QED vacuum) we cannot rely on a definition
to provide the fine tuning, but must perform an actual
calculation. Our model~\cite{QED1} is again a useful tool for a
quantitative analysis.  We have (omitting indices for notational
simplicity; our Bogolubov transformation is diagonal) the following
relation between the squeezing parameter and the Bogolubov coefficient
$\beta$
\begin{equation}
\langle N \rangle = \sinh^2(\zeta) = |\beta|^2.
\end{equation}
By defining $\tau\equiv \pi\; t_0/ ({\textstyle
\ni^2+\no^2})$, where $t_0$ is the timescale on which the refractive
index changes, one has~\cite{QED1}
\begin{eqnarray}
|\beta(\vec k_1,\vec k_2)|^2
&=&
{
\sinh^2\left(
{\textstyle |\ni^2 \omega_\in -\no^2 \omega_\out|} \; \tau
\right)
\over
\sinh\left(
2\; {\textstyle \ni^2 } \; \omega_\in \tau
\right) \;
\sinh\left(
2 \; {\textstyle \no^2} \; \omega_\out \tau
\right)
} \; {V\over(2\pi)^3} \; \delta^3(\vec k_1 + \vec k_2).
\label{beta-squared}
\end{eqnarray}
In the adiabatic limit (large frequencies) we get a Boltzmann
factor~\cite{QED1}
\begin{equation}
|\beta|^2
\approx
\exp\left(-4  \; \min\{\ni,\no\} \no \; \omega_\out \; \tau
\right).
\end{equation}
Since $|\beta|$ is small, $\sinh(\zeta)\approx\tanh(\zeta)$, so that
in this adiabatic limit
\begin{equation}
|\tanh(\zeta)|^2 \approx
\exp\left(-4 \; \min\{\ni,\no\} \no  \; \omega_\out \; \tau
\right).
\end{equation}
Therefore
\begin{equation}
k_{B}T_{\mathrm effective} \approx 
\frac{\hbar}{8\pi t_{0}}\,
\frac{\ni^2+\no^2}{\no \min\{\ni,\no\}}.
\end{equation}
Thus for the entire adiabatic region we can assign a {\em single}
frequency-independent effective temperature, which is really a measure
of the speed with which the refractive index changes. Physically, in
sonoluminescence this observation applies only to the high-frequency
tail of the photon spectrum.

In contrast, in the low frequency region, where the bulk of the
photons emitted in sonoluminescence are to be found, the sudden
approximation holds and the spectrum is phase-space-limited (a power
law spectrum), not Planckian~\cite{QED1}. It is nevertheless still
possible to assign a {\em different} effective temperature for each
frequency.

Finite volume effects smear the momentum-space delta function so we no
longer get exactly back-to-back photons. This represents a further
problem because we have to return to the general squeezed vacuum of
equation (\ref{E:general}). It is still true that photons are emitted
in pairs, pairs that are now approximately back-to-back and of
approximately equal frequency. We can again define an effective
temperature for each photon in the couple as in the ``signal-idler''
systems of quantum optics. This effective temperature is no longer the
same for the two photons belonging to the same couple and no ``special
condition'' for getting the same temperature for all the couples
exists.  Hence the analysis of these finite volume distortions is not
easy~\cite{QED2}, but the qualitative result that in any dynamic
Casimir effect model of sonoluminescence there should be strong
correlations between approximately back-to-back photons is
robust. 

Indeed, if we work with a plane wave approximation for the
electromagnetic eigen-modes (this is essentially a version of the Born
approximation, modified to deal with Bogolubov coefficients instead of
scattering amplitudes) and further modify the infinite-volume model
of~\cite{QED1}, both by permitting a more general temporal profile for
the refractive index, and by cutting off the space integrations at the
surface of the bubble, then the squared Bogolubov coefficient takes
the form
\begin{eqnarray}
|\beta(\vec k_1,\vec k_2)|^2
&=&
F(k_1,k_2; n(t)) \; 
\left| S\left(|\vec k_1 + \vec k_2 |\; R\right) \right|^2.
\label{beta-squared2}
\end{eqnarray}
Here $F(k_1,k_2; n(t))$ is some complicated function of the refractive
index temporal profile, which encodes all the dynamics, while
$S\left(|\vec k_1 + \vec k_2 | \;R\right)$ is a purely kinematical
factor arising from the limited spatial integration:
\[
S\left(|\vec k_1 + \vec k_2 | \;R\right) \equiv 
\int_{r\leq R} d^3\vec r \; \exp\left[-i( \vec k_1 + \vec k_2 ) 
\cdot \vec r \right].
\]
Indeed in the infinite volume limit $|S(\vec k_1,\vec k_2)|^2 \to
[V/(2\pi)^3] \; \delta(\vec k_1+\vec k_2)$.  It is now a standard
calculation to show that
\[
S\left(|\vec k_1 + \vec k_2 | \;R\right) = 
{4\pi\over |\vec k_1 + \vec k_2 |^3} 
\left[ 
\sin(|\vec k_1 + \vec k_2 | \;R) - 
(|\vec k_1 + \vec k_2 | \; R) \;  \cos(|\vec k_1 + \vec k_2 | \; R)
\right].
\]
So, independent of the temporal profile, kinematics will provide
characteristic angular correlations between the outgoing photons:
this result depends only on the the existence of a vacuum squeezing
effect driven by a time-dependent refractive index (which is what is
needed to make the notion of a Bogolubov coefficient meaningful in
this context).  

The plane-wave approximation used to obtain this formula is
valid provided the wavelength of the photons, {\em while they are
still inside the bubble}, are small compared to the dimensions of the
bubble
\begin{equation}
\lambda_{\mathrm{inside}} \ll 
R; \qquad \Rightarrow \qquad \omega \gg {c\over n\; R}.
\end{equation}
While there is still considerable disagreement about the physical size
of the bubble when light emission occurs~\cite{QED0}, and almost no
data concerning the value of the refractive index of the bubble
contents at that time, the scenario developed in~\cite{QED1,QED2} is
very promising in this regard. In particular, high frequency photons
are more likely to exhibit the back-to-back effect, and depending on
the values of $R$ and $n$ this could hold for significant portions of
the resulting emission spectrum. Experimentally, one should work at as
high a frequency as possible---at the peak close to the cutoff.

These observations lead us to the following proposal.

\section{Two-photon observables}

Define the observable
\begin{equation}
N_{ab} \equiv N_{a}-N_{b}, 
\end{equation}
and its variance
\begin{equation}
\Delta (N_{ab})^2=
\Delta N_{a}^{2}+\Delta N_{b}^{2}
-2 \langle N_{a} N_{b}\rangle 
+2 \langle N_{a}\rangle \langle N_{b} \rangle.
\end{equation}
These number operators $N_{a},N_{b}$ are intended to be relative to
photons measured, {\em e.g.}, back to back.  In the case of true
thermal light we get
\begin{equation}
\Delta N_{a}^{2} = \langle N_{a}\rangle(\langle N_{a} \rangle +1),
\end{equation}
\begin{equation}
\langle N_{a} N_{b}\rangle = \langle N_{a}\rangle \langle N_{b}\rangle,
\end{equation} 
so that 
\begin{equation}
\Delta (N_{ab})^2_{\mathrm thermal\ light}
=\langle N_{a}\rangle(\langle N_{a}\rangle+1)
+\langle N_{b}\rangle(\langle N_{b}\rangle+1).
\end{equation}
For a two-mode squeezed-state
\begin{equation}
\Delta (N_{ab})^2_{\mathrm two\ mode\ squeezed\ light}=0.
\end{equation}
Due to correlations, $\langle N_{a} N_{b}\rangle \neq \langle
N_{a}\rangle \langle N_{b}\rangle$. Note also, that if you measure
only a single photon in the couple, you get (as expected) a thermal
variance $\Delta N_{a}^{2} = \langle N_{a}\rangle(\langle N_{a}
\rangle +1) $. Therefore a measurement of the covariance $\Delta
(N_{ab})^2$ can be decisive in discriminating if the photons are
really physically thermal or if non classical correlations
between the photons occur \cite{bk2}.  If the ``thermality'' in the
sonoluminescence spectrum is of this squeezed-mode type, we will
ultimately desire a much more detailed model of the dynamical
Casimir effect involving an interaction term that produces pairs of
photons in two-mode squeezed-states. Apart from our model~\cite{QED1}
and its finite volume generalization~\cite{QED2}, the Eberlein model
also possesses this property~\cite{Eberlein}.  For this type of 
squeezed-mode photon pair-production in a linear medium with
spacetime-dependent dielectric permittivity and magnetic permeability
see \cite{bibi}; for nonlinearity effects see~\cite{lomo}.

In summary: The main experimental signature for squeezed-state photons
being pair-produced in sonoluminescence is the presence of strong
spatial correlations between photons emitted back-to-back and having
the same frequency. These correlations could be measured, for example,
by back-to-back symmetrically placed detectors working in coincidence.
Finite-size effects have been shown in~\cite{QED2} to perturb only
slightly this back-to-back character of the emitted photons, in the
sense that back-to-back emission remains largely dominant.
(Additionally it has been verified that the form of the spectrum is
not violently affected.) Of course, a detailed analysis of the many
technical experimental problems (such as e.g. filtering and multi-mode
signals in the detectors) has also to be done (on these topics
see~\cite{Trentalange}), but such technical details are beyond the
scope of the current work.

\section{Discussion}

The main aims of the present Letter are to clarify the nature of the
photons produced in Casimir-based models of sonoluminescence, and to
delineate the available lines of (theoretical as well experimental)
research that should be followed in order to discriminate
Casimir-based models from thermal models, preferably without having to
understand all of the messy technical details of the condensed matter
physics taking place inside the collapsing bubble.

We have shown that ``effective thermality'' can manifest itself at
different levels.  What is certainly true is that two-mode squeezed
states will exhibit, at a given fixed three-momentum, occupation
numbers which in that mode follow Bose--Einstein statistics. This can
be called ``thermality at fixed wavenumber''.  In contrast, it is
sometimes possible to assign, at least for a reasonably wide range of
wavenumbers, the {\em same} temperature to all modes. This
``thermality across a range of wavenumbers'' gives rise, at least in
this range of wavenumbers, to a spectrum which is approximately
Planckian.

Our sonoluminescence model exhibits Bose--Einstein thermality but not
a truly Planckian spectrum (since the bulk of the photon emission
occurs at frequencies where the sudden approximation holds and a
common temperature for all the momenta is lacking).  The spectrum is
generically a power law at low frequencies followed by a
cut-off~\cite{QED1,QED2}. Although precise measurements in the low
frequency tail of the spectrum could also (in principle) allow us to
discriminate class ``a'' models from class ``b'' models, this
possibility has to be considered strongly model-dependent. Furthermore
the spectral data available at the present time is in this regard
relatively crude: spectral analysis by itself does not seem to be an
appropriate tool for discriminating between class ``a" and class ``b"
models.

Despite this limitation we have shown that there is still the
possibility of obtaining a clear discrimination between real and
effective thermality, without relying on the detailed features of the
model, by looking at two-photon correlations.  For thermal light one
should find thermal variance for photon pairs. On the other hand,
thermofield--like photons should show zero variance in appropriate
pair correlations.  Moreover, our analysis points out that a key point
in discriminating, by means of photon measurements alone, between
classes of models for sonoluminescence is the mechanism of photon
production: Any form of pair-production is associated with two--mode
squeezed states and their strong quantum correlations.  In contrast,
any single-photon production mechanism (thermal, partially thermal,
non-thermal) is not.  In either case, two-photon correlation
measurements are potentially a very useful tool for looking into the
nature of sonoluminescence.

\acknowledgments

This research was supported by the Italian Ministry of Scientific
Research (DWS, SL, and FB), and by the US Department of Energy
(MV). MV particularly wishes to thank SISSA (Trieste, Italy) and
Victoria University (Te Whare Wananga o te Upoko o te Ika a Maui;
Wellington, New Zealand) for hospitality during various stages of this
research. FB is indebted to A.~Gatti for her very helpful remarks
about photon statistics. SL wishes to thank G.~Barton, G.~Plunien, and
R.~\Schutzhold~ for illuminating discussions.

\vskip -0.7 true cm

\end{document}